\documentclass[11pt,twoside]{article}

\usepackage{asp2006_hotcool}
\usepackage{graphicx}
\usepackage{lscape}

\begin{document}
\title{CHARA/MIRC interferometry of red supergiants: diameters, effective 
temperatures and surface features}   
\author{L.L. Kiss\altaffilmark{1}, J.D. Monnier\altaffilmark{2},  
T.R. Bedding\altaffilmark{1}, 
P. Tuthill\altaffilmark{1}, M. Zhao\altaffilmark{2},
M.J. Ireland\altaffilmark{1}, T.A. ten Brummelaar\altaffilmark{3}}
\altaffiltext{1}{Sydney Institute for Astrophysics, School of Physics, University of Sydney,
NSW 2006, Australia}
\altaffiltext{2}{Department of Astronomy, University of Michigan at Ann Arbour, Ann Arbour, MI
48109-1090}
\altaffiltext{3}{The CHARA Array, Georgia State University}

\begin{abstract} 

We have obtained $H$-band interferometric observations of three galactic red supergiant
stars using the MIRC instrument on the CHARA array. The targets include AZ~Cyg, a field RSG
and two members of the Per OB1 association, RS~Per  and T~Per.  We find evidence of
departures from circular symmetry in all cases, which can be modelled with the presence of
hotspots. This is the first detection of these features in the $H$-band. The measured mean
diameters and the spectral energy distributions were combined to determine effective
temperatures. The results give further support to the recently derived hotter temperature
scale of red supergiant stars by Levesque et al. (2005), which has been evoked to reconcile
the empirically determined physical parameters and stellar evolutionary theories. We see a
possible correlation between spottedness and mid-IR emission of the circumstellar dust,
suggesting a connection between mass-loss and the mechanism that generates the spots.

\end{abstract}

\section{Introduction}

Red supergiants (RSGs) represent an important but still poorly characterized
evolutionary phase of massive stars. They are key agents of nucleosynthesis and
chemical evolution of the Galaxy, and have also long been known for their slow
optical variations. This variability is usually attributed to radial pulsations,
although irregular variability caused by huge convection cells was also suggested
from theory (Schwarzschild 1975) and observations of hotspots on late-type
supergiants (Tuthill et al. 1997). From the most extensive study of RSG variability
to date, Kiss et al. (2006) found a strong 1/$f$ noise component in  the
fluctuations of 48 bright galactic objects, a behaviour that fits the picture of
irregular photometric variability caused by large convection cells, analogous to
the granulation background seen in the Sun. In addition, a significant
fraction of the sample shows two distinct time-scales of variability, where the
slower one resembles the enigmatic Long Secondary Periods (LSPs) of the less
massive and luminous Asymptotic Giant Branch stars (Wood et al. 2004).

The best-studied individual RSG in terms of multi-wavelength surface imaging is the
nearest one, Betelgeuse, which has long been known to show a $\sim$400~d pulsation
period and a $\sim$2000~d LSP. For this star, a decade of interferometric hotspot
observations has revealed an irregular shape of the stellar image, the possible
imprint of giant convection cells that were confirmed qualitatively with 3D stellar
convection models (Freytag et al. 2002). Young et al. (2000) found a strong
variation in the apparent asymmetry as a function of wavelength, with a featureless
symmetric disk at 1.290 $\mu$m that is in stark contrast with the presence of
hotspots in the optical range. This has led to the suggestion that the bright spots
are unobscured regions of elevated temperature seen through a
geometrically-extended and line-blanketed atmosphere, in which the features are
seen along  lines of sight for which the atmospheric opacity has been reduced as
the result of activity, such as convection, at the stellar surface (Young et al.
2000). Extending these results to the $H$-band at 1.6 $\mu$m, where an opacity
minimum in the cool atmospheres permits seeing through very close to the
photosphere, one would expect negligible or no evidence of hotspots at this band.

Inspired by the open questions in regards of surface structures of RSGs (e.g.: What
is the physical mechanism that creates the hotspots? How is the spottedness related
to the fundamental stellar parameters and the presence of LSPs?),
we performed a single-epoch mini-survey of several RSGs using the CHARA
interferometric array.

\section{Observations and target selection}

\begin{figure}[!b]
\begin{center}
\includegraphics[width=\textwidth]{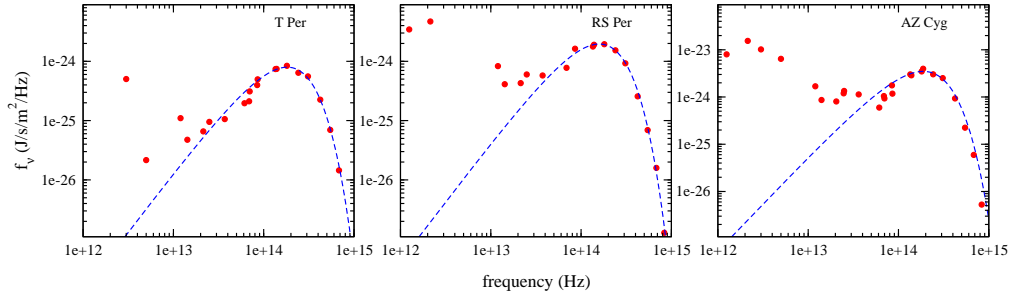}
\caption{Spectral energy distributions of the three observed RSGs. The dashed
lines show blackbody fits to the optical--near-IR parts of the spectra,
unaffected by the circumstellar emission at longer wavelengths.}
\end{center}
\end{figure}

Our observations were carried out on six nights in July-August 2007 at the Georgia
State University Center for High Angular Resolution Astronomy (CHARA)
interferometer array using the Michigan Infra-Red Combiner (MIRC).  The CHARA array,
located on Mount Wilson, consists of six 1 m telescopes. It has
15 baselines ranging from 34 m to 331 m, providing resolutions up to $\sim$0.5 mas
at $H$-band (ten Brummelaar et al. 2005). 

The MIRC instrument was used to combine four CHARA telescopes, providing six
visibilities, four closure phases, and four triple amplitudes simultaneously in
eight narrow spectral bands (Monnier et al. 2006). Using the same W1-W2-S2-E2
configuration of CHARA that was used for surface imaging of Altair (Monnier et al.
2007), we achieved excellent $(u,v)$ coverage of each target. The longest baseline
in this configuration is 251.3 m, corresponding to a resolution of 0.68 mas at
1.6$\mu$m. We observed our targets along with one, two or three calibrators on 
each night, depending on the sky conditions. The data were reduced by the pipeline
outlined by Monnier et al. (2007) and then combined by observing blocks and 
nights.

The initial list of targets contained several stars in the Per OB1 association  and the
galactic field. Of these, we collected data for RS~Per, T~Per,  and AZ~Cygni. Note that
RS~Per is a firmly established member of NGC~884 ($d$=2.34$\pm$0.05 kpc, Slesnick et al.
2002), while T~Per lies about 2 degrees away in the outskirt of the Perseus ``double
cluster''. AZ~Cyg, located in the galactic  plane ($b$=0.52), is not known to be a member
of a young association; in addition to the 495~d pulsation period, it has a $\sim$9-yr LSP
(Kiss et al. 2006).

\begin{figure}[]
\begin{center}
\includegraphics[width=0.7\textwidth]{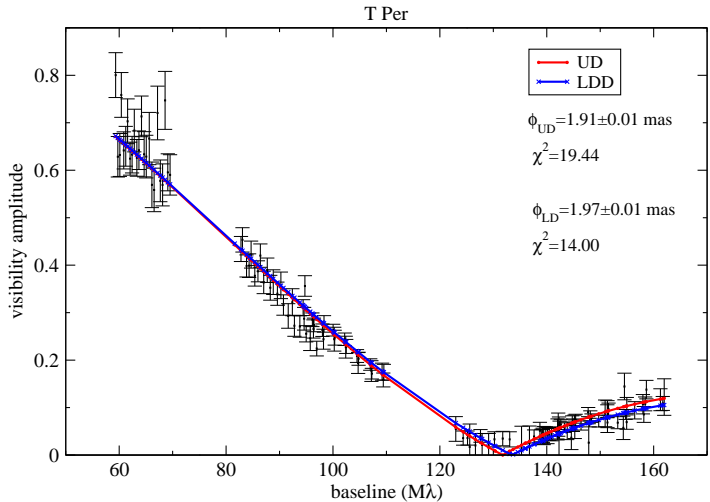}
\caption{Visibility fits for T~Per using uniform and limb-darkened disk models.
The large $\chi^2$ values reflect the simple models' inability to fit the triple
amplitudes and the non-zero closure phases, which indicate significant 
departures from point symmetry.}
\end{center}
\end{figure}

\section{Results}

We have employed different techniques to extract information from the
interferometric data. Simple models were fitted using the latest modification of 
the {\sc mfit}\footnote{\tt http://www.mrao.cam.ac.uk/$\sim$jsy1001/mfit/} 
code by J. Young, originally written by M. Worsley. These included uniform and
limb-darkened disks, elliptical disks and up to two hotspots.
A Hestroffer-type power-law limb darkening with $\alpha$=0.258 
(Lacour et al. 2008) gave remarkably good fits of the visibilities in most cases. 
However, the simplest point-symmetric models fail for each of the three stars 
in closure phases and triple-amplitudes, which leaves no doubt about the detection
of asymmetries in all three stars. We have also obtained preliminary imaging 
using the MACIM code (Ireland et al. 2006).

\subsection{Diameters and effective temperatures}

We have determined the effective temperatures from the limb-darkened angular diameter:

$$T_{\rm eff}=7400\left(\frac{F_{\rm bol}}{10^{-13}~{\rm
Wcm^{-2}}}\right)^{1/4}\left(\frac{1~{\rm mas}}{\phi_{\rm LD}}\right)^{1/2}~{\rm K}$$

\noindent (Perrin et al. 2004). The bolometric flux $F_{\rm bol}$ was computed by integrating a
Planckian fit of the spectral energy distributions (SEDs),  reconstructed from visual ($UBVRI$)
and infrared data ($JHKLM$ + mid- and far-IR fluxes from IRAS, COBE and ISO) taken from 
various catalogues available through the Vizier service of the CDS, Strasbourg. The full SEDs 
(Fig.\ 1) clearly show circumstellar emission in the mid- and far-IR, so that blackbody fits
were restricted to the unaffected shorter-wavelength part of the spectra ($<$5 $\mu$m). 

\begin{figure}[!h]
\begin{center}
\includegraphics[width=0.7\textwidth]{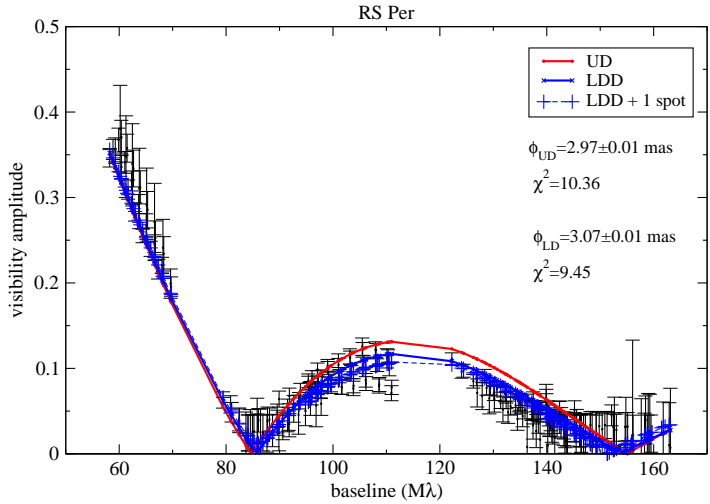}
\caption{Visibility fits for RS~Per using uniform and limb-darkened 
disk models. Adding one hotspot improves the fit to $\chi^2=4.55$ without changing the 
size of the limb-darkened disk.}
\end{center}
\end{figure}

For T~Per and RS~Per, visibility data can be very well fitted with simple models. These are
shown in Figs.\ 2-3. The dereddened fluxes and the LD disk diameters result in the
following effective temperatures: T~Per -- $T_{\rm eff}=3670\pm117$ K; RS~Per -- $T_{\rm
eff}=3551\pm71$ K, where the uncertainties are dominated by the uncertainties in interstellar
reddening. 

\begin{figure}[!h]
\begin{center}
\includegraphics[width=0.75\textwidth]{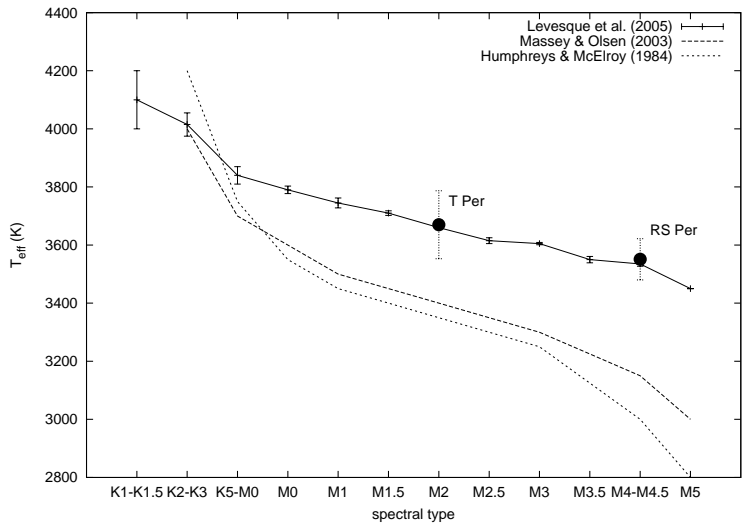}
\caption{Effective temperatures for T~Per and RS~Per, compared to three
temperatures scales from the literature. Our results confirm the hotter scale
of Levesque et al. (2005).}
\end{center}
\end{figure}

The significance of these temperatures is illustrated in Fig.\ 4, where we compare three
RSG effective temperature scales. The excellent agreement between our data and 
the hotter temperature scale of Levesque et al. (2005) gives an independent confirmation of the 
new calibration, which was very successful in reconciling evolutionary 
models and observational properties of red supergiants. 

\subsection{Asymmetries, hotspots}

We see evidence for hotspots with $\sim$10\% flux contribution in all three stars, implying
that: {\it (i)} hotspots are indeed  ubiquitous in red supergiants; and {\it (ii)} these
structures are visible in $H$-band too, in contrast to the Young et al. (2000) results for
Betelgeuse at 1.29 $\mu$m. Considering the opacity minimum at 1.6 $\mu$m, these hotspots
must be generated very close to the photosphere and their enhanced contrast can hardly be
explained by opacity effects. Furthermore, in our admittedly small sample, the complexity
of the observed stars seems to increase with the strength of circumstellar IR emission,
which may indicate a connection between the hotspots and mass loss.

The most complex structures have been revealed for AZ~Cyg, where the data suggest a 
brightness distribution almost like a contact binary with a $\sim$10-20\% bright  feature
at the NW edge of the disk. Simple LDD + 1 spot models (Fig.\ 5) are roughly consistent
with MACIM fits, but the analysis of this star is still in progress.

\begin{figure}[!h]
\begin{center}
\includegraphics[width=0.73\textwidth]{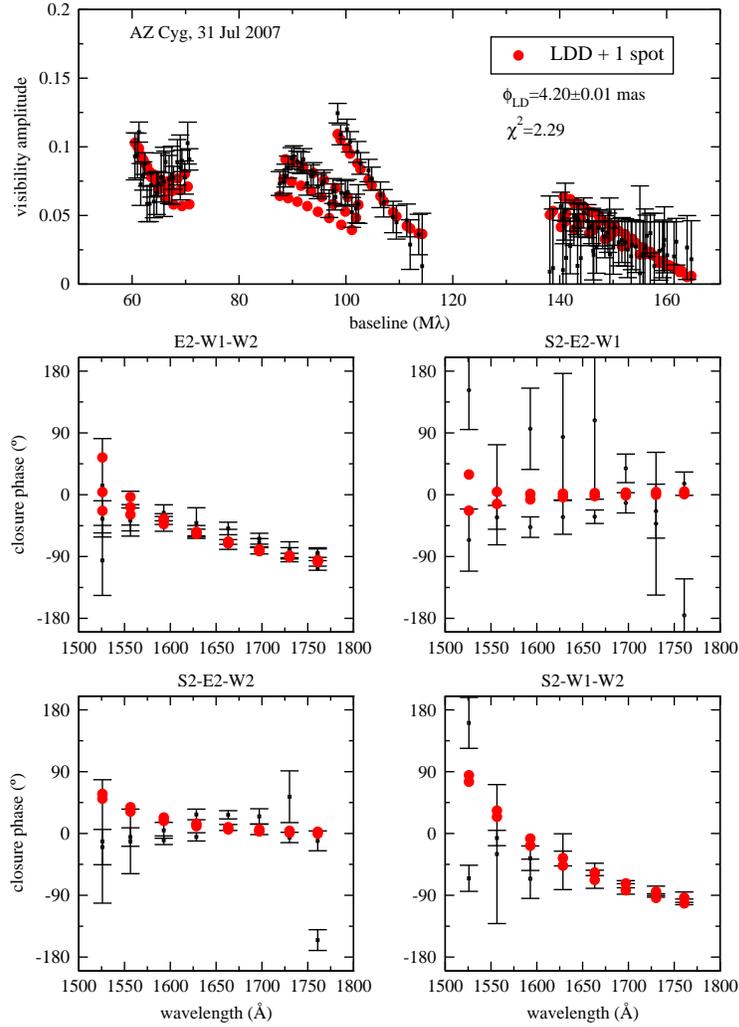}
\caption{Visibilities and closure phases for AZ~Cyg on 31 July, 2007.}
\end{center}
\end{figure}

\acknowledgements 

This project has been supported by the Australian Research Council. 


\end{document}